\documentclass[a4paper]{article}

\usepackage{INTERSPEECH2021}
\usepackage{subfigure}
\title{VAENAR-TTS: Variational Auto-Encoder based Non-AutoRegressive Text-to-Speech Synthesis}
\name{Hui Lu$^{1,2}$, Zhiyong Wu$^{1,3}$, Xixin Wu$^4$, Xu Li$^1$, Shiyin Kang$^5$, Xunying Liu$^1$, Helen Meng$^{1,2,3}$}
\address{
$^1$Dept. of Systems Engineering \& Engineering Management,
Chinese University of Hong Kong\\
$^2$Centre for Perceptual and Interactive Intelligence, CUHK\\
$^3$Tsinghua-CUHK Joint Research Center for Media Sciences, Technologies and Systems,\\
Shenzhen International Graduate School, Tsinghua University, Shenzhen, China\\
$^4$Department of Engineering, University of Cambridge, UK \\
$^5$Huya Inc., Guangzhou, China}
\email{\{luhui, zywu, wuxx, xuli, xyliu, hmmeng\}@se.cuhk.edu.hk, kangshiyin@huya.com}

\begin{document}
\setlength{\belowdisplayskip}{4pt} \setlength{\belowdisplayshortskip}{4pt}
\setlength{\abovedisplayskip}{4pt} \setlength{\abovedisplayshortskip}{4pt}

\maketitle
\begin{abstract}
This paper describes a variational auto-encoder based non-autoregressive text-to-speech (VAENAR-TTS) model. The autoregressive TTS (AR-TTS) models based on the sequence-to-sequence architecture can generate high-quality speech, but their sequential decoding process can be time-consuming. Recently, non-autoregressive TTS (NAR-TTS) models have been shown to be more efficient with the parallel decoding process. However, these NAR-TTS models rely on phoneme-level durations to generate a hard alignment between the text and the spectrogram. Obtaining duration labels, either through forced alignment or knowledge distillation, is cumbersome. Furthermore, hard alignment based on phoneme expansion can degrade the naturalness of the synthesized speech. In contrast, the proposed model of VAENAR-TTS is an end-to-end approach that does not require phoneme-level durations. The VAENAR-TTS model does not contain recurrent structures and is completely non-autoregressive in both the training and inference phases. Based on the VAE architecture, the alignment information is encoded in the latent variable, and attention-based soft alignment between the text and the latent variable is used in the decoder to reconstruct the spectrogram. Experiments show that VAENAR-TTS achieves state-of-the-art synthesis quality, while the synthesis speed is comparable with other NAR-TTS models.
\end{abstract}

\section{Introduction}

Text-to-speech (TTS) synthesis aims to generate speech from input text. With the help of deep learning methods, TTS models are able to produce high-quality and natural-sounding speech comparable with that uttered by a human. A modern TTS system mainly involves two steps: spectrogram prediction from the input text and waveform reconstruction from the spectrogram. Focusing on the former step, we propose a VAE-based, end-to-end NAR-TTS model that yields high speech quality with fast synthesis speed.


AR-TTS models \cite{tacotron2017,shen2018natural,li2019neural, ping2018deep} utilize the attention mechanism \cite{seq2seqattention} to align and predict the next spectrogram frame based on the currently generated ones. This paradigm relaxes the need for explicit duration modeling in traditional TTS models \cite{ze2013statistical}, and can potentially enhance the naturalness of the synthesized speech. However, the time complexity of the decoding process grows linearly with the length of the spectrogram. In addition, the error may accumulate throughout the sequential generation process, which affects especially synthesis for long sentences.


Phoneme-level durations have been used in many NAR-TTS models to eliminate the AR decoding process. Phoneme-level durations can be distilled from a pre-trained AR-TTS model \cite{lee2021bidirectional,fastspeech2019}, extracted via force-alignment \cite{fastspeech2} or from dynamic programming obtained alignments \cite{glowtts2020}. Based on the acquired durations, these models expand the linguistic feature into the frame-level feature, which can be further mapped to the spectrogram non-autoregressively. Without AR decoding, the synthesis speed is significantly improved. However, deriving phoneme-level durations
can be cumbersome. Also, the hard-alignment of the linguistic feature to the frame-level by repeating phonemes may degrade the naturalness of the synthesized speech. Previous work \cite{fastspeech2} attempted to alleviate such degradation by incorporating greater variations in pitch and energy, but this also increases the complexity of model training.



Other NAR-TTS models utilize utterance-level duration, that is, the number of frames of an utterance, to create a placeholder and use it to build up the spectrogram in a NAR way \cite{paranet2020,flowtts2020}.
The prediction of the utterance-level duration is more convenient than that of phoneme-level durations as the utterance-level duration is inherently available. However, it can be challenging to obtain the alignment between the linguistic feature and the spectrogram placeholder. Though positional encoding of the utterance-level duration can be used to assist in learning the alignment, it does not provide sufficient information to encode the linguistic pattern inside the spectrogram. The pressure in learning the alignment can be further mitigated by introducing an AR-TTS teacher \cite{paranet2020}, but this also makes the training process more complicated.



This paper proposes a novel approach for TTS that is NAR and based on VAE, denoted as VAENAR-TTS. In contrast to NAR-TTS models based on phoneme-level durations, VAENAR-TTS requires only the text-spectrogram pair and thus can avoid the complexities of the forced alignment or knowledge distillation processes as mentioned above. Hence, VAENAR-TTS offers greater simplicity and is more straightforwardly end-to-end. This is achieved by using the VAE architecture to encode the alignment into a latent variable. The prior and posterior distributions of this alignment-aware latent variable is learned in the VAE training paradigm. During the inference phase, the latent variable is sampled from the prior distribution conditioned on the linguistic feature. The latent variable encodes both the linguistic and alignment information , and can be used as the spectrogram placeholder to be aligned with the linguistic feature. This attention-based soft alignment of the linguistic feature is more suitable for natural speech synthesis. The sample shape of the latent variable is determined by the utterance-level duration predicted from the linguistic feature.

The architecture of VAENAR-TTS is mainly inspired by a NAR machine translation (MT) model \cite{flowseq2019}. However, since the task of TTS differs significantly from MT, we have made multiple modifications to adapt it across the tasks.
The VAENAR-TTS model use Glow \cite{kingma2018glow} to model the prior distribution of the latent variable, as compared with Glow-based NAR-TTS models \cite{flowtts2020,glowtts2020} that adopt it as the decoder to predict the spectrogram. Previous work has used bidirectional hierarchical VAE to model NAR-TTS \cite{lee2021bidirectional}, but in this work, we use only the vanilla conditional VAE structure.


\section{Model Formalization}
\label{sec:background}

TTS can be formalized as the modeling of the posterior probability distribution of the spectrogram $Y=[y_1,y_2,...,y_N]$ given the linguistic feature $X=[x_1,x_2,...,x_M]$, denoted as $P(Y|X)$, where $N$ and $M$ are the number of frames of the spectrogram and the number of characters in the text, respectively. $y_i$ is the $i^{th}$ frame of the spectrogram and $x_j$ is the linguistic vector corresponding to the $j^{th}$ character in the text, where $i=1,...,N; j=1,...,M$. In AR-TTS models, $P(Y|X)$ is factorized as follows:
\begin{equation}
P(Y|X)=\prod_{i=1}^{N}{P(y_i|y_{-i}, X)}.
\label{equ:ar-tts-prob}
\end{equation}
This factorization yields very good results since the spectrogram is continuous and thus it is easy to infer a feature frame from the preceding frames. But the autoregressive decoding process can be time-consuming and vulnerable to accumulated error. We adopt another form of factorization as shown below:
\begin{equation}
P(Y|Z,X)=\prod_{i=1}^{N}{P(y_i|Z,X)}.
\label{equ:nar-tts-prob}
\end{equation}
By introducing the latent variable $Z$, each frame of the spectrogram becomes independent of one another given the linguistic feature $X$ and $Z$, thus can be generated in parallel. For TTS, there are multiple kinds of information that can be encoded in $Z$ to satisfy this factorization. For example, phoneme-level durations can be used as $Z$ to extend the phoneme-level linguistic feature to the frame-level, then the spectrogram can be predicted non-autoregressively with a frame-to-frame mapping model. This factorization with $Z$ to model phoneme-level durations has been implicitly adopted in previous work \cite{fastspeech2019,fastspeech2} and yields very good performance. However, obtaining phoneme-level duration labels requires much extra effort and the hard alignment between the linguistic feature and the spectrogram may cause unnaturalness of the synthesized speech. To avoid these issues, the proposed approach does not restrict the latent variable in modeling specific piece(s) of expert knowledge. Instead, the latent variable is learned by the model itself in a variational inference paradigm.

The computation of the posterior probability distribution $P(Y|X)$ based on Equation (\ref{equ:nar-tts-prob}) can be approximated with its evidence lower bound shown in Equation (\ref{equ:nar-elbo}), where $P(Z|X)$ is the prior distribution of $Z$ given the linguistic feature, $Q(Z|X,Y)$ is the posterior distribution of $Z$ given the linguistic feature and the ground-truth spectrogram. The first term in Equation (\ref{equ:nar-elbo}) is the negative Kullback–Leibler divergence (KL-divergence) between the posterior distribution $Q(Z|X,Y)$ and the prior distribution $P(Z|X)$. The second term denotes the expected log-likelihood of the spectrogram.
\begin{align}
\log[Pr(Y|X)]\approx &-\int_{Z}{Q(Z|X,Y)\log[\frac{Q(Z|X,Y)}{P(Z|X)}]dZ} \nonumber \\
&+ \int_{Z}{Q(Z|X,Y)\log[P(Y|Z,X)]dZ}
\label{equ:nar-elbo}
\end{align}
With the above formalization, the architecture of the proposed VAENAR-TTS can be readily derived.
\section{Architecture}
\label{sec:architecture}
\begin{figure}[t]
\centering
\includegraphics[width=0.9\linewidth]{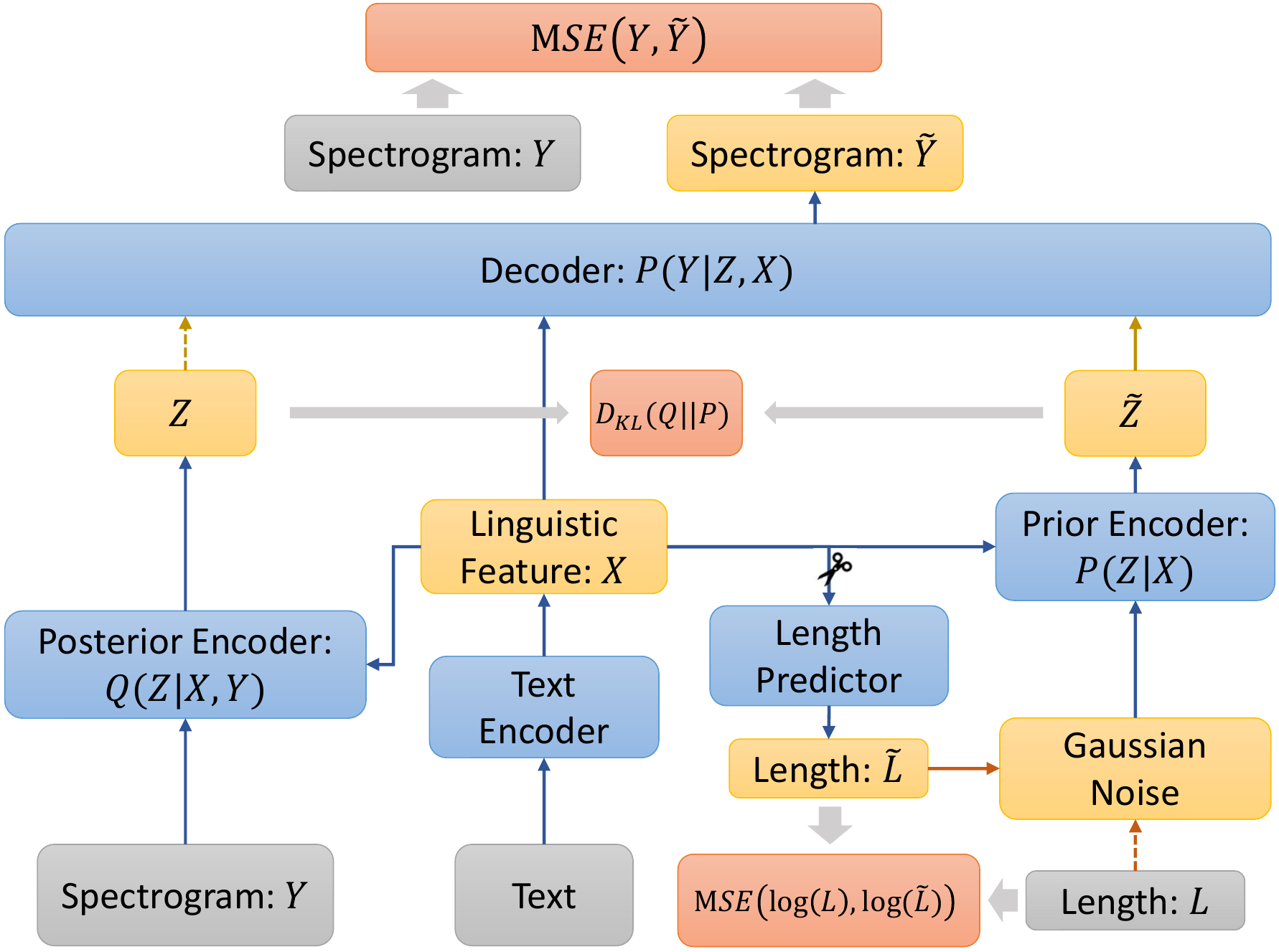}
\caption{Architecture of VAENAR-TTS. The dotted lines are only turned on during the training stage, while their counterparts with the same color are only turned on during the inference phase. The scissors denote that the gradients for the length predictor will not be back-propagated to the text encoder. }
\label{fig:model_architecture}
\end{figure}

The architecture of VAENAR-TTS is shown in Figure \ref{fig:model_architecture}. It consists of a text encoder, a posterior encoder, a prior encoder, a length predictor, and a decoder. The text encoder aims to encode the raw character sequence into the context-aware linguistic feature $X$. The prior encoder models the prior distribution of $Z$ conditioned on $X$: $P(Z|X)$; while the posterior encoder models the posterior distribution of $Z$ given the spectrogram $Y$ and $X$: $Q(Z|X,Y)$.
The length predictor is built to infer the utterance-level duration from the linguistic feature. The posterior can be more informative about the alignment since it is conditioned on the ground-truth spectrogram. The prior is pushed towards the posterior by the KL-divergence loss during the training phase. During the inference phase, the latent variable with the predicted length is sampled from the prior distribution. With the sampled latent variable as the base, the decoder aligns the linguistic feature onto it and reconstructs the spectrogram, which corresponds to $P(Y|Z,X)$ in Equation (\ref{equ:nar-elbo}).

Conditioning on the linguistic feature is accomplished through the attention mechanism, to which the linguistic feature is used as the key and value being queried. Though many neural architectures with attention mechanisms can serve this need, we adopt self-attention blocks and decoder attention blocks from Transformer \cite{transformer2017} as the main components of our model. The architecture design details are described as follows.

\textbf{Text Encoder}: The text encoder adopts similar structure as that in Transformer-TTS \cite{li2019neural}. It consists of a convolutional PreNet that includes a convolution stack with dropout \cite{JMLR:v15:srivastava14a}, batch normalization \cite{pmlr-v37-ioffe15} and ReLU activation \cite{nair2010rectified}. The sinusoidal positional encoding \cite{transformer2017} is added to keep the temporal information. Then, several self-attention blocks are stacked upon to model the contextual information in the text.

\textbf{Posterior Encoder}: In the posterior encoder, the spectrogram is first fed into a dense PreNet which consists of two fully-connected layers with dropout and ReLU activation. The sinusoidal positional encoding is added as well. With the encoded spectrogram as the query and the linguistic feature as the key and value, several attention blocks are stacked upon the PreNet to compute the alignment. Following the Transformer decoder, the attention block consists of a self-attention layer, a cross-attention layer, and an FFN layer. The frame-level mean and variance for the posterior distribution of the latent variable $Z$ is predicted based on the outputs of the attention blocks.

\textbf{Prior Encoder}: Following \cite{flowseq2019}, we use a Glow \cite{kingma2018glow} structure to model the prior distribution of the latent variable. The prior encoder consists of multiple Glow blocks with each block containing an actnorm layer, an invertible $1\times1$ convolution followed by an affine-coupling layer. The transformation neural network in the affine-coupling layer is based on the Transformer decoder block, with the latent variable as the query and the linguistic feature as the key and value. While the latent variable $Z$ is sampled via the forward pass of the prior, the backward pass can be used to infer the probability of a specific latent variable.

\textbf{Decoder}: The decoder consists of several Transformer decoder blocks to align the linguistic feature with the latent variable. The spectrogram is then predicted from the aligned linguistic feature. The convolution-based PostNet \cite{shen2018natural} is used to generate a residual complement for the predicted spectrogram.

\textbf{Length Predictor}: The length predictor is just a 1-channel fully-connected layer followed by the ReLU activation, whose character-level outputs can be considered as the logarithm character-level durations. The utterance-level duration is obtained by summing up the exponential of the character-level outputs. The utterance-level duration cannot be accurately predicted since the lengths of pause or silence segments at the beginning and ending part of an utterance are nontrivial to estimate. However, we find that adding a constant bias to the predicted duration can prevent the synthesized speech from being uncompleted while not degrading the speech quality, which only appends a short segment of silence to the synthesized speech.

\textbf{Loss Function}: The training loss function of VAENAR-TTS is shown below:
\begin{align}
L = &\mathbf{MSE}(Y, \tilde{Y}) + \alpha \mathbf{D_{KL}}(Q(Z|X,Y)||P(Z|X)) \nonumber\\
&+ \beta \mathbf{MSE}(\log(L), \log(\tilde{L})),
\label{equ:trainingloss}
\end{align}
where $\mathbf{MSE}$ denotes the mean squared error, $\mathbf{D_{KL}}$ is the KL-divergence. $Y$, $\tilde{Y}$, $L$ and $\tilde{L}$ are the ground-truth and the predicted spectrogram, the ground-truth and the predicted utterance-level duration, respectively. $\alpha$ and $\beta$ are two weight hyper-parameters. Note that we use the logarithm of the utterance-level duration to compute the MSE as the loss term for the length predictor.

\section{Alignment Learning}
\label{sec:alignment_learning}

The accurate alignment between the linguistic feature and the spectrogram is the key to high-quality speech synthesis. Previous AR-TTS models use location-sensitive attention \cite{lsa2015} to help learn a monotonic alignment trajectory, which is an AR process. We utilize two other methods to help learn the accurate alignment without introducing extra AR components.


\textbf{Annealing Reduction Factor}: Intuitively, sequences of shorter lengths can be more easily aligned. However, the spectrogram of a typical utterance usually consists of hundreds of frames, which increases the difficulty of the alignment learning, especially in the early training stage.
The reduction factor $r$ has been used to control the number of frames output at each decoder step \cite{tacotron2017}. We use a similar technique in the proposed model. While a larger $r$ enables faster alignment convergence, it also degrades the speech quality since the reduced spectrogram may lose some fine-grained information that is important for learning an accurate posterior distribution of the latent variable. To alleviate this issue, we adopt the annealing reduction factor strategy in the training of VAENAR-TTS. To be specific, the initial $r$ is set to a relatively large value to facilitate the learning of diagonal attention alignments.
After the nearly diagonal alignments is acquired, $r$ is decreased gradually to enable the learning of the fine-grained distributions and spectrogram.

\textbf{Causality Mask}: The causality mask is added to the self-attention stacked on the frame-level feature, namely, the self-attention in the posterior encoder, the prior encoder, and the decoder. The causality mask restricts the attention region to the current frame and the frames before the current frame. We find that restricting the self-attention on the frame-level feature to be causal can help reduce the repetition errors as it strengthens the temporal information in the frame-level feature and helps learn better alignment.

\section{Experiments}
\label{sec:exp}

\subsection{Experimental Setup}
\label{sec:exp:hparams}

The experiments are conducted on LJSpeech \cite{ljspeech17} which contains 13,100 English utterances from a female speaker. Two 131-utterance subsets are randomly sampled out as the validation and test set while the remaining are the training set. This dataset separation configuration is shared among all of the compared models.

We use the 80-dimension logarithm Mel-spectrogram which is extracted with the 256-sample window shift and 1024-sample window length as the prediction target. For multi-head attention in all attention structures, the output dimension is set to 256, the number of heads is 4, the FFN hidden dimension is set to 1024. The number of attention blocks in the text encoder, posterior, decoder, and each prior Glow block are 4, 2, 2, and 2, respectively. For the text encoder, 43 lower-case letters and symbols are embedded into 512-dimension. The convolutional PreNet in the text encoder contains 5 layers of 1D convolution with 512 filters and the kernel size of 5. The dense PreNet in the posterior consists of 2 fully connected layers with 256-dimension outputs. The prior encoder contains 6 Glow blocks. The dimension of the latent variable is 128. The model is trained on an RTX2080Ti GPU with the batch size of 32. The learning rate is constantly set to $1.25\times10^{-4}$. The optimizer is Adam \cite{adam2015} with $\beta_1=0.9$ and $\beta_2=0.999$. The weights for KL-divergence and the utterance-level duration loss are set to $1.0\times10^{-5}$ and $1.0$ respectively. The reduction factor annealing strategy is as follows: $r$ is initially set to 5 and is decreased by 1 every 200 training epochs until it reaches 2, after which $r$ remains as 2 for the rest of the training epochs. We train the model for 2000 epochs and the final model checkpoint is used for evaluation. During the training phase, the initial noise for the prior encoder is sampled from the normal distribution, while for inference it is set to all zeros, we find that the synthesized speech is more stable in this way. A constant bias of 80 frames is added to the predicted utterance-level duration to avoid the synthesized utterance being incomplete during inference.

We compare the proposed model with the state-of-the-art AR-TTS model: Tacotron2 \cite{shen2018natural}, and NAR-TTS models: FastSpeech2 \cite{fastspeech2}, Glow-TTS \cite{glowtts2020}, and BVAE-TTS \cite{lee2021bidirectional}. We use Hifi-GAN \cite{hifigan2020} to convert the generated spectrogram into the waveform for all of the compared models. To analyze the effect of the reduction factor $r$, we also evaluate 3 VAENAR-TTS models with $r$ fixed for the whole training process as 5, 4, and 3, denoted as RF5, RF4, and RF3, respectively.

\subsection{Synthesis Quality and Speed Experiments}
\label{sec:exp:exp_results}
10 out of 131 sentences from the test set are randomly selected and synthesized by each model as the subjective evaluation set\footnote{Samples and code: https://github.com/thuhcsi/VAENAR-TTS}. We compare different TTS models in terms of synthesis quality as well as synthesis speed. The synthesis quality is measured by the subjective evaluation in terms of the Mean Opinion Score (MOS) of the speech naturalness. 15 subjects are invited to evaluate the naturalness of the synthesized speech on a 5-scale basis in which 1 means bad and 5 means excellent naturalness. The synthesis speed is measured using the real-time factor (RTF), which is the time taken to synthesize one second of the speech spectrogram from the text. The speed tests are conducted on a single RTX2080Ti GPU with the batch size of 1. For each model, the RTF is averaged over 10 runs on the whole test set.

The results are shown in Table \ref{tab:quality_speed_results}. The MOS scores are presented with $95\%$ confidence intervals. We can observe that VAENAR-TTS achieves the best speech naturalness among all the compared models with MOS of $4.15\pm0.13$, while is comparable or better than Tacotron2 with MOS of $4.03\pm0.12$. As for the synthesis speed, the proposed model can synthesize 1 second of speech spectrogram within $7.45$ milliseconds, which is comparable with other NAR-TTS models, and is about $18\times$ faster than Tacotron2 .
\begin{table}[t]
\caption{Comparison results of different TTS models}
\label{tab:quality_speed_results}
\centering
\begin{tabular}{lll}
\toprule
\textbf{Model} & \textbf{MOS} & \textbf{RTF}(Sec) \\
\midrule
Ground-Truth & $4.56\pm0.09$ & - \\
Hifi-GAN-Resyn & $4.47\pm0.10$ & - \\
Tacotron2 & $4.03\pm0.12$ & $1.35\times10^{-1}$ \\
FastSpeech2 & $3.83\pm0.14$ & $\mathbf{4.21\times10^{-3}}$ \\
Glow-TTS & $3.62\pm0.13$ & $9.39\times10^{-3}$ \\
BVAE-TTS & $3.16\pm0.13$ & $\mathbf{4.21\times10^{-3}}$ \\
VAENAR-TTS & $\mathbf{4.15\pm0.12}$ & $7.45\times10^{-3}$ \\
\midrule
RF5 & $3.43\pm0.14$ & $6.99\times10^{-3}$ \\
RF4 & $3.83\pm0.13$ & $7.30\times10^{-3}$ \\
RF3 & $3.84\pm0.14$ & $7.43\times10^{-3}$ \\
\bottomrule
\end{tabular}
\end{table}
\subsection{Alignment Learning Experiments}
\label{sec:exp:exp_results}
\begin{figure}[t]
\centering
\includegraphics[width=0.8\linewidth]{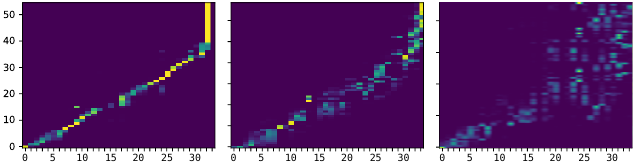}
\caption{Attention alignments of RF5 model (left), RF4 model (middle) and RF3 model (right) after 56 training epochs}
\label{fig:rf543_ali}
\end{figure}

The lower part of Table \ref{tab:quality_speed_results} shows the experimental results of VAENAR-TTS with different fixed reduction factors $r$. We can see a significant improvement of speech naturalness when $r$ is decreased from 5 to 4, while the MOS gap between RF4 and RF3 is relatively small. With the final $r$ as 2, VAENAR-TTS with annealing $r$ achieves much better quality than RF3. The MOS results also support our argument that as the reduction factor increases, more fine-grained information may be lost, which hinders the estimation of the latent distributions, thus degrades the naturalness of the predicted speech. It is also worth noting that RF3 and RF4 models achieve better MOS scores than Glow-TTS and BVAE-TTS while is comparable with FastSpeech2. Though the frame-level feature is further reduced, as compared with VAENAR-TTS, there are no significant drops of RTFs for these models since the spectrogram is generated in parallel.
Besides, the decoder attention alignments of models with larger $r$ converge much faster. As we observe in Figure \ref{fig:rf543_ali}, after 56 epochs of training, RF5 can already obtain a relatively clear diagonal alignment trajectory, while the attention maps for RF4 and RF3 are rather blurred. Further training shows that RF4 acquires clear diagonal alignment after training epoch 96 while this number for RF3 is about 122.
\begin{figure}[t]
\centering
\includegraphics[width=0.52\linewidth]{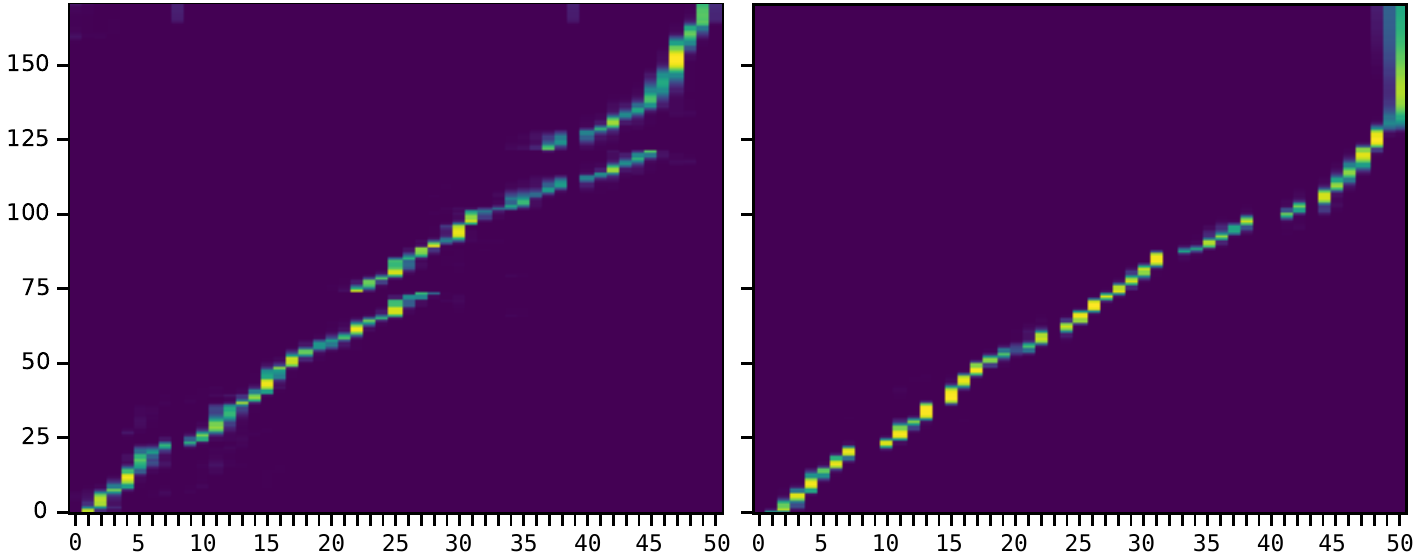}
\caption{Decoding attention alignments of models without (left) versus with (right) causality mask in acoustic side self-attention, where the vertical and horizontal axis denotes the decoder and encoder step, respectively.}
\label{fig:causalmask}
\end{figure}

VAENAR-TTS model without the causality mask in the frame-level feature side self-attention is also trained and compared. But the repetition problems come frequently to this ablation model and the MOS score cannot reflect this issue more than the attention alignment shown in Figure \ref{fig:causalmask}. As we can see, though the attention alignment is clear, without the causality mask, the attention trajectory goes back several times to attend to the characters that are already synthesized, this causes the repetition problem in the synthesized speech. With the causality mask added, the self-attention computation can add more temporal information to the frame-level feature, which we argue is very important for the continuous and monotonic alignment learning as shown in the right part of Figure \ref{fig:causalmask}.

\section{Conclusions}
\label{sec:conclusions}
We present a VAE-based NAR-TTS model named VAENAR-TTS in this paper. Compared with recently proposed NAR-TTS models, VAENAR-TTS requires no phoneme-level durations and thus avoids the complex duration extraction process. VAENAR-TTS utilizes VAE to encode the alignment information into the latent variable, and uses the attention-based soft alignment to align the linguistic feature with the frame-level latent variable. In this way, our approach can alleviate the unnaturalness problem caused by the hard alignment based on phoneme-level durations. We adopt the annealing reduction factor and causality mask in self-attention to help learn the alignment faster and better. VAENAR-TTS achieves state-of-the-art synthesis quality as well as competitive synthesis speed to that of other NAR-TTS models.
\hfill \vskip 5pt
\noindent
\textbf{Acknowledgements}: This work is partially supported by National Key R\&D Program of China (2020AAA0104500), National Natural Science Foundation of China (NSFC) (62076144), the Major Project of National Social Science Foundation of China (NSSF) (13\&ZD189), and is partially supported by the Centre for Perceptual and Interactive Intelligence, a CUHK InnoCentre.

\bibliographystyle{IEEEtran}

\bibliography{mybib}

\begin{thebibliography}{10}
\providecommand{\url}[1]{#1}
\csname url@samestyle\endcsname
\providecommand{\newblock}{\relax}
\providecommand{\bibinfo}[2]{#2}
\providecommand{\BIBentrySTDinterwordspacing}{\spaceskip=0pt\relax}
\providecommand{\BIBentryALTinterwordstretchfactor}{4}
\providecommand{\BIBentryALTinterwordspacing}{\spaceskip=\fontdimen2\font plus
\BIBentryALTinterwordstretchfactor\fontdimen3\font minus
  \fontdimen4\font\relax}
\providecommand{\BIBforeignlanguage}[2]{{%
\expandafter\ifx\csname l@#1\endcsname\relax
\typeout{** WARNING: IEEEtran.bst: No hyphenation pattern has been}%
\typeout{** loaded for the language `#1'. Using the pattern for}%
\typeout{** the default language instead.}%
\else
\language=\csname l@#1\endcsname
\fi
#2}}
\providecommand{\BIBdecl}{\relax}
\BIBdecl

\bibitem{tacotron2017}
Y.~Wang, R.~J. Skerry{-}Ryan, D.~Stanton, Y.~Wu, R.~J. Weiss, N.~Jaitly,
  Z.~Yang, Y.~Xiao, Z.~Chen, S.~Bengio, Q.~V. Le, Y.~Agiomyrgiannakis,
  R.~Clark, and R.~A. Saurous, ``Tacotron: Towards end-to-end speech
  synthesis,'' in \emph{Interspeech 2017, 18th Annual Conference of the
  International Speech Communication Association, Stockholm, Sweden, August
  20-24, 2017}, F.~Lacerda, Ed.\hskip 1em plus 0.5em minus 0.4em\relax {ISCA},
  2017, pp. 4006--4010.

\bibitem{shen2018natural}
J.~Shen, R.~Pang, R.~J. Weiss, M.~Schuster, N.~Jaitly, Z.~Yang, Z.~Chen,
  Y.~Zhang, Y.~Wang, R.~Skerrv-Ryan \emph{et~al.}, ``Natural tts synthesis by
  conditioning wavenet on mel spectrogram predictions,'' in \emph{2018 IEEE
  International Conference on Acoustics, Speech and Signal Processing
  (ICASSP)}.\hskip 1em plus 0.5em minus 0.4em\relax IEEE, 2018, pp. 4779--4783.

\bibitem{li2019neural}
N.~Li, S.~Liu, Y.~Liu, S.~Zhao, and M.~Liu, ``Neural speech synthesis with
  transformer network,'' in \emph{Proceedings of the AAAI Conference on
  Artificial Intelligence}, vol.~33, no.~01, 2019, pp. 6706--6713.

\bibitem{ping2018deep}
W.~Ping, K.~Peng, A.~Gibiansky, S.~O. Arik, A.~Kannan, S.~Narang, J.~Raiman,
  and J.~Miller, ``Deep voice 3: 2000-speaker neural text-to-speech,''
  \emph{Proc. ICLR}, pp. 214--217, 2018.

\bibitem{seq2seqattention}
D.~Bahdanau, K.~Cho, and Y.~Bengio, ``Neural machine translation by jointly
  learning to align and translate,'' in \emph{3rd International Conference on
  Learning Representations, {ICLR} 2015, San Diego, CA, USA, May 7-9, 2015,
  Conference Track Proceedings}, Y.~Bengio and Y.~LeCun, Eds., 2015.

\bibitem{ze2013statistical}
H.~Ze, A.~Senior, and M.~Schuster, ``Statistical parametric speech synthesis
  using deep neural networks,'' in \emph{2013 ieee international conference on
  acoustics, speech and signal processing}.\hskip 1em plus 0.5em minus
  0.4em\relax IEEE, 2013, pp. 7962--7966.

\bibitem{lee2021bidirectional}
Y.~Lee, J.~Shin, and K.~Jung, ``Bidirectional variational inference for
  non-autoregressive text-to-speech,'' in \emph{International Conference on
  Learning Representations}, 2021.

\bibitem{fastspeech2019}
Y.~Ren, Y.~Ruan, X.~Tan, T.~Qin, S.~Zhao, Z.~Zhao, and T.~Liu, ``Fastspeech:
  Fast, robust and controllable text to speech,'' in \emph{Advances in Neural
  Information Processing Systems 32: Annual Conference on Neural Information
  Processing Systems 2019, NeurIPS 2019, December 8-14, 2019, Vancouver, BC,
  Canada}, H.~M. Wallach, H.~Larochelle, A.~Beygelzimer,
  F.~d'Alch{\'{e}}{-}Buc, E.~B. Fox, and R.~Garnett, Eds., 2019, pp.
  3165--3174.

\bibitem{fastspeech2}
Y.~Ren, C.~Hu, X.~Tan, T.~Qin, S.~Zhao, Z.~Zhao, and T.~Liu, ``Fastspeech 2:
  Fast and high-quality end-to-end text to speech,'' \emph{CoRR}, vol.
  abs/2006.04558, 2020.

\bibitem{glowtts2020}
J.~Kim, S.~Kim, J.~Kong, and S.~Yoon, ``Glow-tts: {A} generative flow for
  text-to-speech via monotonic alignment search,'' in \emph{Advances in Neural
  Information Processing Systems 33: Annual Conference on Neural Information
  Processing Systems 2020, NeurIPS 2020, December 6-12, 2020, virtual},
  H.~Larochelle, M.~Ranzato, R.~Hadsell, M.~Balcan, and H.~Lin, Eds., 2020.

\bibitem{paranet2020}
K.~Peng, W.~Ping, Z.~Song, and K.~Zhao, ``Non-autoregressive neural
  text-to-speech,'' in \emph{Proceedings of the 37th International Conference
  on Machine Learning, {ICML} 2020, 13-18 July 2020, Virtual Event}, ser.
  Proceedings of Machine Learning Research, vol. 119.\hskip 1em plus 0.5em
  minus 0.4em\relax {PMLR}, 2020, pp. 7586--7598.

\bibitem{flowtts2020}
C.~Miao, S.~Liang, M.~Chen, J.~Ma, S.~Wang, and J.~Xiao, ``Flow-tts: {A}
  non-autoregressive network for text to speech based on flow,'' in \emph{2020
  {IEEE} International Conference on Acoustics, Speech and Signal Processing,
  {ICASSP} 2020, Barcelona, Spain, May 4-8, 2020}.\hskip 1em plus 0.5em minus
  0.4em\relax {IEEE}, 2020, pp. 7209--7213.

\bibitem{flowseq2019}
X.~Ma, C.~Zhou, X.~Li, G.~Neubig, and E.~Hovy, ``Flowseq: Non-autoregressive
  conditional sequence generation with generative flow,'' in \emph{Proceedings
  of the 2019 Conference on Empirical Methods in Natural Language Processing},
  Hong Kong, November 2019.

\bibitem{kingma2018glow}
D.~P. Kingma and P.~Dhariwal, ``Glow: generative flow with invertible 1$\times$
  1 convolutions,'' in \emph{Proceedings of the 32nd International Conference
  on Neural Information Processing Systems}, 2018, pp. 10\,236--10\,245.

\bibitem{transformer2017}
A.~Vaswani, N.~Shazeer, N.~Parmar, J.~Uszkoreit, L.~Jones, A.~N. Gomez,
  L.~Kaiser, and I.~Polosukhin, ``Attention is all you need,'' in
  \emph{Advances in Neural Information Processing Systems 30: Annual Conference
  on Neural Information Processing Systems 2017, December 4-9, 2017, Long
  Beach, CA, {USA}}, I.~Guyon, U.~von Luxburg, S.~Bengio, H.~M. Wallach,
  R.~Fergus, S.~V.~N. Vishwanathan, and R.~Garnett, Eds., 2017, pp. 5998--6008.

\bibitem{JMLR:v15:srivastava14a}
N.~Srivastava, G.~Hinton, A.~Krizhevsky, I.~Sutskever, and R.~Salakhutdinov,
  ``Dropout: A simple way to prevent neural networks from overfitting,''
  \emph{Journal of Machine Learning Research}, vol.~15, no.~56, pp. 1929--1958,
  2014.

\bibitem{pmlr-v37-ioffe15}
S.~Ioffe and C.~Szegedy, ``Batch normalization: Accelerating deep network
  training by reducing internal covariate shift,'' in \emph{Proceedings of the
  32nd International Conference on Machine Learning}, ser. Proceedings of
  Machine Learning Research, F.~Bach and D.~Blei, Eds., vol.~37.\hskip 1em plus
  0.5em minus 0.4em\relax Lille, France: PMLR, 07--09 Jul 2015, pp. 448--456.

\bibitem{nair2010rectified}
V.~Nair and G.~E. Hinton, ``Rectified linear units improve restricted boltzmann
  machines,'' in \emph{Icml}, 2010.

\bibitem{lsa2015}
J.~Chorowski, D.~Bahdanau, D.~Serdyuk, K.~Cho, and Y.~Bengio, ``Attention-based
  models for speech recognition,'' in \emph{Advances in Neural Information
  Processing Systems 28: Annual Conference on Neural Information Processing
  Systems 2015, December 7-12, 2015, Montreal, Quebec, Canada}, C.~Cortes,
  N.~D. Lawrence, D.~D. Lee, M.~Sugiyama, and R.~Garnett, Eds., 2015, pp.
  577--585.

\bibitem{ljspeech17}
K.~Ito and L.~Johnson, ``The lj speech dataset,''
  \url{https://keithito.com/LJ-Speech-Dataset/}, 2017.

\bibitem{adam2015}
D.~P. Kingma and J.~Ba, ``Adam: {A} method for stochastic optimization,'' in
  \emph{3rd International Conference on Learning Representations, {ICLR} 2015,
  San Diego, CA, USA, May 7-9, 2015, Conference Track Proceedings}, Y.~Bengio
  and Y.~LeCun, Eds., 2015.

\bibitem{hifigan2020}
J.~Kong, J.~Kim, and J.~Bae, ``Hifi-gan: Generative adversarial networks for
  efficient and high fidelity speech synthesis,'' in \emph{Advances in Neural
  Information Processing Systems 33: Annual Conference on Neural Information
  Processing Systems 2020, NeurIPS 2020, December 6-12, 2020, virtual},
  H.~Larochelle, M.~Ranzato, R.~Hadsell, M.~Balcan, and H.~Lin, Eds., 2020.

\end{thebibliography}

\end{document}